\documentclass[floatfix,twocolumn,aps,prl,amsmath,amssymb]{revtex4}
\usepackage[latin1]{inputenc}
\usepackage[dvips]{graphicx}
\usepackage{dcolumn}
\usepackage{bm}
\usepackage{color}
\usepackage{url}
\usepackage[colorlinks=true ,citecolor=blue]{hyperref}
\usepackage{amsmath,amssymb}

\begin{document}

\title{The Resistivity of High-Tc Cuprates}
\author{R. Arouca$^{1,2}$}\thanks{arouca@pos.if.ufrj.br}
\author{E. C. Marino$^{1}$}\thanks{marino@if.ufrj.br}
\affiliation{$^1$Instituto de F\' isica, Universidade Federal do Rio de Janeiro, C.P. 68528, Rio de Janeiro, RJ, 21941-972, Brazil.}%
\affiliation{$^2$Institute for Theoretical Physics, Center for Extreme Matter and Emergent Phenomena, Utrecht University, Princetonplein 5, 3584 CC Utrecht, The Netherlands.}

\date{\today}

\begin{abstract}
 We show that the resistivity in each phase of the High-Tc cuprates is a special case of a general expression derived from the Kubo formula. We obtain, in particular, the T-linear behavior in the strange metal (SM) and upper pseudogap (PG) phases, the pure $T^2$, Fermi liquid (FL) behavior observed in the strongly overdoped regime as well as the $T^{1+\delta}$ behavior that interpolates both in the crossover. We calculate the coefficients: a) of $T$ in the linear regime and show that it is proportional to the PG temperature $T^*(x)$; b) of the $T^2$-term in the FL regime, without adjusting any parameter; and c) of the $T^{1.6}$ term in the crossover regime, all in excellent agreement with the experimental data. From our model, we are able to infer that the resistivity in cuprates is caused by the scattering of holes by excitons, which naturally form as holes are doped into the electron background.

\end{abstract}
\maketitle

{\bf Introduction} High-$T_c$ superconductivity in the cuprates  \cite{bednorz1986possible} is, at the same time, one of the most interesting and complex problems in contemporary physics. Although some features of these systems, such as the relevance of magnetic interactions in the $CuO_2$ planes \cite{dagotto94,scalapino12} and the non s-wave  character of the superconducting order parameter \cite{scalapino86, wollman93, tsuei00}, are consensual in the community, there are important issues that, so far, remain unsolved.

	Among the fundamental unanswered questions related to the cuprates, there are some that concern the normal state of these materials. A particularly intriguing one is: how to explain the perfectly linear dependence of the resistivity with the temperature \cite{hu2017universal, legros2019universal}, which is observed in all cuprate materials?  This deviates from the well-known $T^2$ behavior, typical of Fermi liquids, which are usually associated with conventional metals \cite{marino2017quantum,mahan2013many,coleman2015introduction}. Nevertheless, the metallic nature of this state is suggested by the resistivity increase with the temperature, hence justifying the name strange metal (SM), by which this phase is known \cite{gurvitch87, keimer2015quantum}.
	 In the attempt to explain the linear behavior of the resistivity in this phase, different mechanisms have been proposed \cite{varma89, varma99, faulkner2010strange, zaanen14, Sachdev18}. Among these, we find the ``Planckian dissipation hypothesis" \cite{zaanen2004superconductivity, legros2019universal, zaanen19planckian} that associates the scattering rate of the charge carriers, $1/\tau$, with the inverse characteristic time of thermal fluctuations: $k_B T/\hbar$, which ultimately follows from the uncertainty principle. 
	 Also the state responsible for the linear resistivity in cuprates has been associated to a regime of quantum criticality, namely,  a quantum critical point (QCP) would exist, producing a phase consisting in a quantum critical fluid whose properties would be universally determined\cite{sachdev97, sachdev_2011}. 

Nevertheless, as discussed in Ref.~\cite{phillips05}, the Planckian dissipation hypothesis (PDH) combined with scaling arguments would be, in principle, not compatible with a T-linear dependence of the resistivity. Yet, it is argued in Ref.~\cite{phillips05} that in case resistivity could be ascribed to the scattering of holes out of a bosonic field, then the PDH in a quantum critical regime could be reconciled with the linear behavior of resistivity.
	 
	 The QCP supposedly responsible for this universal quantum critical behavior has been associated to a metal-insulator quantum phase transition shown to exist at a doping value $x_p$ located inside the SC dome of these materials under the effect of strong magnetic fields that would destroy the SC state \cite{keimer2015quantum,taillefer2010scattering, vishik2012phase}. The assumption that the PG temperature transition line $T^*(x)$ ends at a point $x_p$, which is inside the SC dome however, does not seem to have experimental support, since no other $T^*(x)$ points are observed inside the dome \cite{vishik2012phase}.

In a recent publication, we proposed a model which provides a general and accurate description of the high-Tc superconductivity in cuprates. This model allowed for the obtainment of analytical expressions for the SC and PG temperatures: $T_c(x)$ and $T^*(x)$, showing excellent agreement with the experimental data for different compounds \cite{marino2020superconducting}. This analytical solution clearly shows that the PG temperature line meets the SC temperature line at $T=0$, on a QCP located at the right extremity of the SC dome.


The model we propose for understanding the cuprates \cite{marino2020superconducting}, exhibits two quartic interaction terms: a) one which is hole-attractive and derives from the magnetic Kondo interaction between the itinerant holes and localized copper spins; and b) another, which is hole-repulsive and stems from the Hubbard electric repulsion between the holes.
 Performing a Hubbard-Stratonovitch (HS) transformation in both terms, we introduce two HS scalar fields, respectively, $\Phi$ and $\chi$. $\Phi$ is the creation operator of Cooper pairs, that condense on the Superconducting (SC) phase. $\chi$, by its turn, is the creation operator of excitons, (electron-hole bound-states) that, upon condensation, give rise to the Pseudogap (PG) phase \cite{marino2020superconducting}. With the aid of this picture, we are able to conclude that the main mechanism responsible for resistivity in the normal phases of cuprates is the scattering of charged holes by excitons. Such exciton states should be observable in the insulating regime of the cuprates, namely, in the low-temperature region of the strongly underdoped PG phase.
 
  	In the present study, we shall insert the current correlators derived from our model into the Kubo formula at a finite temperature, in order to obtain a general expression for the resistivity, which reduces to the ones occurring in the different normal states of the High-Tc cuprates.

{\bf The Model and Resistivity} We take the model derived in  Ref.~\cite{marino2020superconducting} as the starting point.
After performing a HS transformation with the scalar fields $\Phi$ and $\chi$ in each of the two quartic terms, our Hamiltonian becomes  \cite{marino2020superconducting}
	\begin{eqnarray}
		&\ &H_{eff}=\sum_{\textbf{k},\sigma}\epsilon(\textbf{k}) \Big [\psi_{A\sigma}^\dagger(\textbf{k})\psi_{B\sigma}(\textbf{k})+hc\Big ]
		\nonumber \\
		&\ &+ \sum_{\textbf{k}}\Phi(\textbf{k}) \Big [\psi_{A\uparrow}^\dagger(-\textbf{k})
		\psi^\dagger_{B\downarrow}(\textbf{k}) + 
		\psi^\dagger_{B\uparrow}(\textbf{k}) \psi_{A\downarrow}^\dagger(-\textbf{k})\Big]+hc
		\nonumber \\
		&\ &
		+\sum_{\textbf{k}}\chi(\textbf{k}) \Big [\psi_{A\sigma}^\dagger(\textbf{k})
		\psi_{B\sigma}(\textbf{k})  \Big ]+ hc
		\nonumber \\
		&\ &
		+ \frac{1}{g_S}\sum_{\textbf{k}}\Phi^\dagger(\textbf{k}) \Phi(\textbf{k})+ \frac{1}{g_P}\sum_{\textbf{k}}\chi^\dagger(\textbf{k}) \chi(\textbf{k}),
		\label{eq_Ham}
	\end{eqnarray}
where $\psi_{a\sigma}$ represents a hole on the oxygen sublattices $a=A,B$ with spin $\sigma=\uparrow,\downarrow$. $g_S$ is the SC pairing coupling parameter while $g_P$ is the coupling responsible for the PG phase transition. In the above expression $\epsilon(\textbf{k})$ is the usual tight-binding kinetic energy of the free-holes on a square lattice. Notice that the PG field $\chi$ acts as a scattering potential for the holes, being therefore responsible for their resistivity, whereas the SC field $\Phi$ is related to the formation of Cooper pairs.

	Integrating on the fermionic (holes) degrees of freedom, we arrive at an effective thermodynamic potential that depends  on the SC and PG order parameters, respectively, $\Delta=\langle\Phi\rangle$ and $M=\langle\chi\rangle$, as well as on the chemical potential $\mu$: $\Omega(\Delta, M,\mu,T)$ (See Supplemental Material).  Using that potential, we could explain the SC phase diagram of different cuprate compounds, obtaining in particular, the left and right limiting points of the SC dome, namely $x_{SC}^{-}$ and $x_{SC}^{+}\equiv\tilde{x}_0$ \cite{marino2020superconducting}.

In order to obtain the resistivity in our model, we first introduce an external electromagnetic field, through the minimal coupling of the kinetic term with the vector potential $\textbf{A}$: $\epsilon(\textbf{k}) \rightarrow \epsilon(\textbf{k}+ e \textbf{A}).$ The Hamiltonian, then becomes $H \rightarrow H[\textbf{A}]$, and out of this, we 
obtain the grand-partition functional $Z[\textbf{A}]$
which yields the grand-canonical potential in the presence of an applied electromagnetic vector potential $ \textbf{A}$, namely, $\Omega [ \textbf{A} ]$.

The average electric current and its correlation functions are, then, obtained from the expressions:
\begin{eqnarray}
\langle j^i \rangle = \frac{\delta  \Omega [ \textbf{A} ] }{\delta \textbf{A}^i}\ \ \ \ ,\langle j^i j^j\rangle = \frac{\delta^2  \Omega [ \textbf{A} ] }{\delta \textbf{A}^i \delta \textbf{A}^j}.
\label{eq_j}
\end{eqnarray}



 We, then use the Kubo formula at a finite temperature \cite{mahan2013many, marino2017quantum}, and the
 current-current correlation function obtained from the 
  grand-canonical potential  $\Omega$, derived from our model, Eq.~\eqref{eq_j}, in order to obtain an explicit expression for the conductivity per $CuO_2$ plane. Upon inversion, this leads to a general expression for the resistivity per plane in the normal phase ($\Delta=0$) (Supplemental Material)
	\begin{align}
	\begin{split}
		\rho\left(x,T\right)=&\frac{Vk_B}{\hbar v^2e^2}T \frac{M\left[\cosh\left(\frac{M}{k_BT}\right)+\cosh\left(\frac{\mu}{k_BT}\right)\right]}{\sinh\left(\frac{M}{k_BT}\right)}
		\label{eq_rho}
	\end{split}
	\end{align}
	or
	\begin{equation}
\rho(x,T)=BT^2 G\left(\frac{M}{k_B T}, \frac{\mu}{k_B T}\right),
\label{r}
\end{equation}

where $V=da^2$ is the volume of the primitive unit cell, per $CuO_2$ plane, $h/e^2\approx 25 812.807 \Omega$ is the resistance quantum, $d$ is the distance between planes, $a$ is the lattice parameter and $v$ is the characteristic velocity of the holes, such that $\left(\hbar v/a\right)\approx 2.9\times 10^{-2} eV$~\cite{marino2020superconducting}. The resistivity is expressed in terms of the constant
	\begin{equation}
		B=\frac{h}{e^2}\frac{d}{2\pi}\left(\frac{a}{\hbar v}\right)^2k_B^2\approx 3.62\ d\times  10^{-4}\  \mu \Omega \text{cm}/K^2,
		\label{eq_B}
	\end{equation}
for $d$ in \AA\ -units, and $G(K_1, K_2)$, the scaling function of the critical variables $K_1=\frac{M}{k_B T}$ and $K_2=\frac{\mu}{k_B T}$, given by
	\begin{equation}
		G\left(K_1,K_2\right)=K_1\frac{\cosh K_1 +\cosh K_2}{2\sinh \left(K_1\right)}.
		\label{eq_G}
	\end{equation}

	This general form of the resistivity holds in all phases of the phase diagram of cuprates, except the SC one. The peculiar
	form of the resistivity in each of the different phases will be determined by the specific form of the function                      $G\left(K_1,K_2\right)$ in each phase.


	\begin{figure}[!h]
		\centering
		\includegraphics[width=\linewidth]{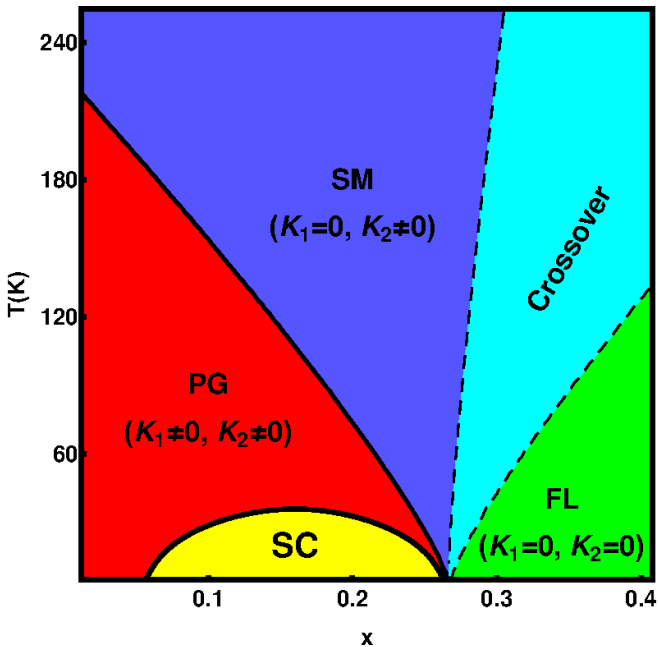}
		\caption{Phase diagram of LSCO (not a cartoon!) shows the general features of the phase diagram of hole-doped High-Tc Cuprates. The solid lines are our theoretical expressions of $T^*(x)$ and $T_c(x)$ derived in Ref.~\cite{marino2020superconducting}, while the dashed lines are the curves $T'^*(x)\equiv T^*\left(2\tilde{x}_0-x\right)$ and $T_{\text{cross}}=C_0T'^*(x)$ (with $C_0$ defined below) that roughly defines the crossover between the FL and SM phases. The phase diagram for other High-Tc cuprate compounds is similar apart from the asymmetry, which is observed between the overdoped and underdoped regions in the SC dome. The magnetically ordered phase displayed by all cuprates in the weakly doping regime has been studied elsewhere \cite{marino01} and is omitted here. }
		\label{fig_phase_diagram}
	\end{figure}

{\bf The Cuprates Phase Diagram}	As discussed thoroughly in \cite{marino2020superconducting}, we can understand the phase diagram of cuprates by studying the behavior of the grand-canonical potential as a function of $\Delta$ and $M$ as well as of the temperature and chemical potential. Analytical expressions for, $T_c\left(x\right)$ and $T^*\left(x\right)$, were obtained, which set the boundaries of the different phases for doping parameter $x$ below the critical point at $\tilde{x}_0$, where all phase boundaries meet. For doping larger than $\tilde{x}_0$ we have the FL phase, at sufficiently low temperatures, the SM phase at high temperatures and a crossover between the two \cite{nakamae03, cooper2009anomalous}. The LSCO phase diagram, displaying the analytical expressions for the SC and PG temperatures: $T_c\left(x\right)$ and $T^*\left(x\right)$, as well as the crossover temperatures: $T'^*(x)\equiv T^*(2\tilde{x}_0-x)$ and $T_{\text{cross}}=C_0T'^*(x)$ (with $C_0$ defined below) is presented in Fig~\ref{fig_phase_diagram}. We can classify the normal state phase diagram in terms of the variables $K_1$ and $K_2$. 

{\bf Resistivity in the PG Phase} In this phase, both $K_1$ and $K_2$ are different from zero.
We subdivide the PG phase in three regions according to the values of $x$, namely: a) strongly underdoped, for $x<x_{SC}^{-}$; b) underdoped, for $x_{SC}^{-}<x\lesssim x_0$, where $x_0$ is the optimal doping; c) overdoped, for $x_0\lesssim x<x_{SC}^{+}$.

{\bf Resistivity in the PG Phase: High Temperature Regime} In the three subregions, we first consider the temperature range close to $T^*(x)$, where $M\rightarrow 0$, implying $K_1\rightarrow 0$. Then (see Supplemental Material)
		\begin{equation}
			G_{PG}( T\rightarrow T^*) =\frac{T^*}{T}\cosh^2\left(\frac{\mu_{PG}(T\rightarrow T^*)}{2 k_B T}\right),  
			\label{eq_G_PG_Ts}
		\end{equation}
		Inserting this in (\ref{eq_rho}) we obtain for the resistivity
		
		\begin{equation}
			\rho_{PG}( T\rightarrow T^*) =B T^*T\cosh^2\left(\frac{\mu_{PG}( T\rightarrow T^*)}{2 k_B T}\right),  
			\label{eq_rho_PG_Ts}
		\end{equation}
{\bf Resistivity in the PG Phase: Low Temperature Regime} In all subregions, except for the strongly underdoped region, the low-temperature regime is dominated by the SC phase, where the resistivity vanishes.
 In order to explore the resistivity in the strongly underdoped region, we take the low-temperature limit of (\ref{eq_G}), obtaining
		\begin{eqnarray}
			G_{PG}( T\rightarrow 0)&\approx &K_1 \exp\left(K_1-K_2\right)\\ \nonumber
			&=&\frac{M}{k_B T} \exp\left(\frac{\mu-M}{k_B T}\right),  
			\label{eq_G_PG_0}
		\end{eqnarray}
which makes that the resistivity in this limit to be given by
		\begin{equation}
			\rho_{PG} \left(x, T\rightarrow 0\right)=\frac{B M}{2 k_B} T \exp\left(\frac{T_0}{T}\right),  
			\label{eq_rho_PG_0}
		\end{equation}
such that it presents an insulating behavior with an activation temperature $T_0$ given by $\frac{\mu\left(T,x\right)-M\left(T,x\right)}{k_B}$.
	
{\bf Resistivity on the SM Phase} In the SM phase, $M=0$ and $K_1=0$ for all temperatures, so that the function $G$ is given by
	\begin{equation}
		G_{SM}=\lim_{M\rightarrow 0} \frac{M}{T\sinh\left(\frac{M}{2 k_B T}\right)}\cosh^2\left(\frac{\mu_{SM}}{2 k_B T}\right).  
		\label{eq_G_SM_1}
	\end{equation}
The SM phase corresponds to the quantum critical region associated to the QCP located at the right end of the Sc dome, namely $x_{SC}^{+}\equiv \tilde{x}_0$. It follows that, in that region, all quantities with dimension of energy should scale with $T$ \cite{phillips2012advanced, kirkpatrick15}. In particular, the chemical potential must be given by
	\begin{equation}
		\mu_{SM}\left(T,x\right)=D k_B T,
	\end{equation}
where $D$ is a constant \cite{sachdev_2011, phillips2012advanced}. In this way, we see that in the SM phase $K_2=D$ is constant. Imposing the continuity of the resistivity and its derivative across the border between the PG and SM phases, at $T=T^*$ implies that the scaling function of the SM phase is given by
	\begin{equation}
		G_{SM}=G_{PG}( T\rightarrow T^*)= \frac{C T^*}{T},  
		\label{eq_G_SM_2}
	\end{equation}
where $C=\cosh^2\left(\frac{D}{2}\right)$. This immediately yields the celebrated linear resistivity
		\begin{align}
		\begin{split}
			\rho_{SM}=A_1 T = \left( C B  T^*\right) T
			\label{eq_rho_SM}
		\end{split}
		\end{align}
which ranges from the upper PG phase all the way into the SM phase. For regions of the SM phase where $x>x_{SC}^{+}\equiv \tilde{x}_0$, the temperature $T^*(x)$ is smoothly replaced by $T'^*(x)\equiv T^*(2\tilde{x}_0-x)$

{\bf Resistivity in the FL Phase} In the FL phase, which corresponds to the strongly overdoped regime, $x>x_{SC}^{+}\equiv \tilde{x}_0$, at low temperatures, both $K_1$ and $K_2$ are equal to zero. Consequently (see Supplemental Material)
	\begin{equation}
		G_{FL}=G\left(K_1=0, K_2=0\right)=1,
		\label{eq_G_FL}
	\end{equation} 
and then, in this phase, the resistivity is given by the quadratic behavior, typical of a Fermi liquid:
	\begin{equation}
		\rho_{FL}=B T^2,
		\label{eq_rho_FL}
	\end{equation}
with the coefficient $B$ given by (\ref{eq_B}).

\begin{figure*}[!hbt]
	\centering
	\includegraphics[width=0.75\linewidth]{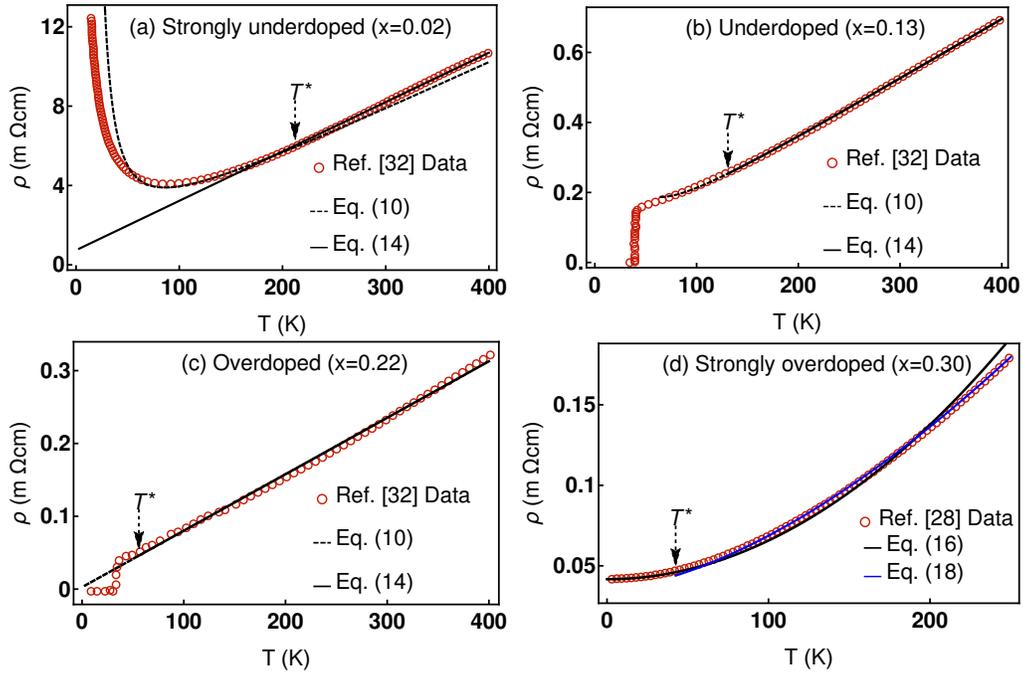}
	\caption{Evolution of the resistivity behavior as a function of doping for LSCO. (a) Strongly underdoped regime ($x<x_{\text{SC}}^{-}$):  there is no SC phase and $T_0<T^*$. The resistivity diverges for temperatures below $T_0$. (b) Underdoped regime ($x_{\text{SC}}^{-}<x\lesssim x_0$): $T_0$ diminishes and the upturn of resistivity occurs for lower temperatures and appears as a kink close to $T_c$. (c) Overdoped regime ($x_0\lesssim x< x_{\text{SC}}^{+}$: $T_c$ gets higher than $T_0$ and the resistivity becomes completely linear in the normal phase. (d) Strongly overdoped regime ($x>x_{\text{SC}}^{+}$): for low enough temperatures, the resistivity has the quadratic behavior typical of the FL phase with the coefficient given by Eq.~\eqref{eq_B}, while for higher temperature it has the power law behavior of Eq.~\eqref{eq_rho_cross}. The experimental data shown above are for LSCO compounds with doping levels of $x=0.02$ (strongly underdoped), $x=0.13$ (underdoped), $x=0.22$ (overdoped) and $x=0.3$ (strongly overdoped). The red circles are data extracted from Ref.~\cite{ando2004electronic} and Ref.~\cite{nakamae03}. The values of the coefficient of the FL and crossover phases are obtained by Eq.~\eqref{eq_B} and Eq.~\eqref{eq_rho_cross} and not by a fitting process.}
	\label{fig_LSCO_res}
\end{figure*}

	{\bf Crossover} The transition between FL and the SM phases is not really, a phase transition, but rather a crossover with an intermediate power-law behavior $\rho\sim T^{1+\delta}$, $0\leq\delta\leq 1$, that interpolates the resistivity behaviors, namely, linear and quadratic, of the SM and FL phases \cite{nakamae03, cooper2009anomalous}. This can be accounted for assuming that $G$ has the power-law behavior
	\begin{equation}
		G_{\text{cross}}=\left[\frac{C_0 T'^*(x)}{T}\right]^{1-\delta},
		\label{eq_G_cross}
	\end{equation}
	which is obtained from the previous one (\ref{eq_rho_SM}) by a scale transformation.
The resistivity, then, will be given by
	\begin{equation}
		\rho_{\text{cross}}=B\left[C_0 T'^*(x)\right]^{1-\delta}T^{1+\delta}\equiv B_C T^{1+\delta}.
		\label{eq_rho_cross}
	\end{equation}
for $\delta \in [0,1]$. Notice that the values $\delta =0$ and $\delta =1$, respectively, correspond to the SM and FL phases, which are, thereby interpolated by the above expression. Observe, also, that the specific function interpolating the resistivity expressions in the SM and FL phases could be continuously deformed, but keeping its form at the limiting values of $\delta$ similarly to what happens to the members of the same homotopy class.
The crossover behavior occurs in the region of the phase diagram located between $T'^*(x)$ and the temperature $T_{\text{cross}}\equiv C_0 T'^*(x) $.

\begin{figure*}[!htb]
	\centering
	\includegraphics[width=\linewidth]{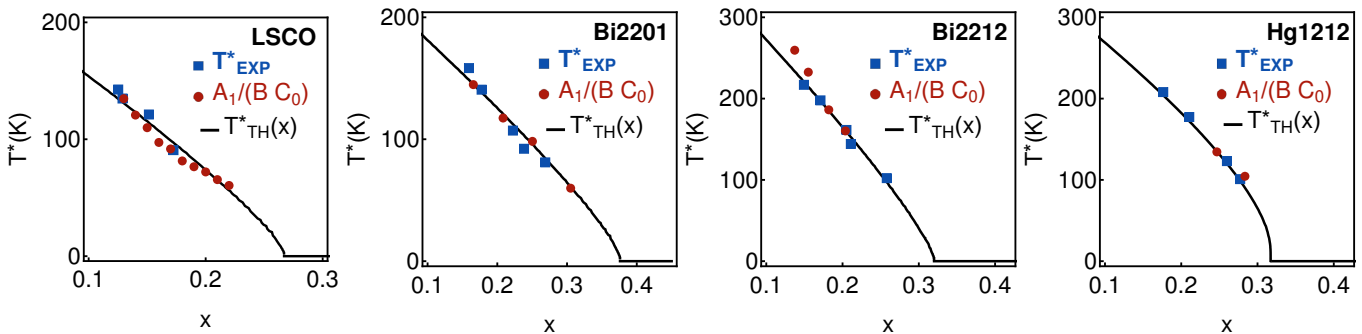}
	\caption{Scaling of $A_1/(BC_0)$ (red circles) with $T^*$, both with our theoretical expression (black solid line) as well as experimental values (blue squares), for different compounds.~The experimental data was extracted from: LSCO ~\cite{ando2004electronic}, Bi2201 ~\cite{ando2000carrier}, Bi2212 ~\cite{akoshima98} and Hg1212 ~\cite{yamamotopress}. The values of $C_0$ and $d$ used are in Table~\ref{tab_C}.}
	\label{fig_alpha_Ts}
\end{figure*}

{\bf Comparison with Experimental Data} We have successfully applied the above theoretical framework in the description of the resistivity of cuprates. Our results accurately explain the experimental data of several compounds, namely: LSCO ~\cite{ando2004electronic, cooper2009anomalous, nakamae03}, Bi-based (Bi2201 ~\cite{ando2000carrier}, Bi2212 ~\cite{akoshima98}) and Hg-based (Hg1212 ~\cite{yamamotopress}) families of cuprates. Our strategy was to fit $A_1$ in the metallic regimes and $T_0$, specifically  in the insulating state occurring in the low-temperature, strongly underdoped regime of the PG phase.
 The details of the methodology as well as all analyses are presented on the Supplemental Material.
	
	In order to determine the value of $C$ we must consider the ratio $A_1(x)/\left[B(d) T^*(x)\right]$, for different values of the doping parameter $x$. As it turns out, as we increase $x$ thus moving towards the quantum critical point, this stabilizes at a constant value $C_0$. For LSCO, this occurs for $0.10 \lesssim x$.
	 The constant behavior of $C$ implies that in the quantum critical region, the $x$-dependence of the resistivity slope comes through the dependence of $T^*$ on $x$. In other words, $A_1$ scales with the PG temperature $T^*(x)$.
A similar behavior of $A_1$ has been reported in the literature in Ref.~\cite{taillefer2010scattering}, even though it was not associated to the PG temperature.

\begin{figure}[!htb]
	\centering
	\includegraphics[width=\linewidth]{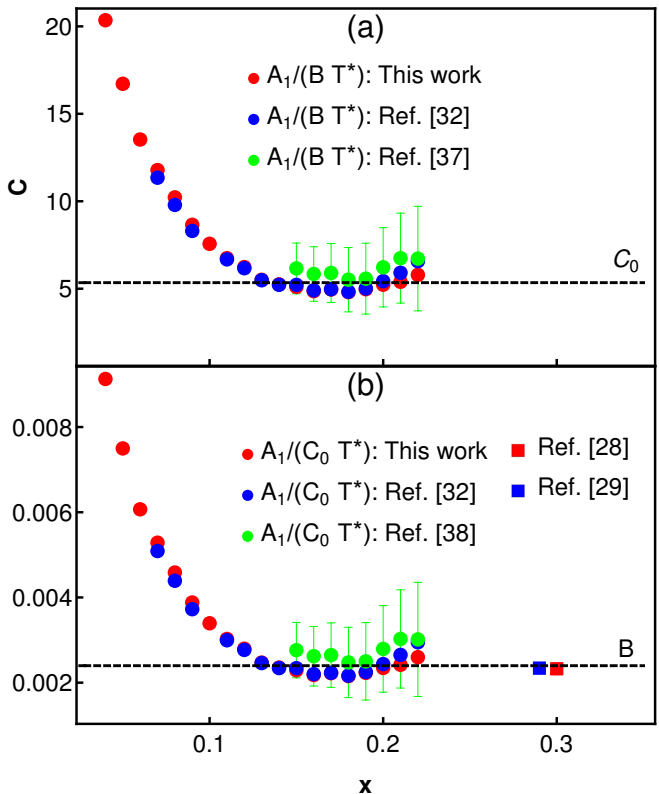}
	\caption{(a) Evolution of $A_1/(B T^*)$ with doping, showing that it stabilizes in the value $C_0$ (black dashed line) displayed in Tab.~\ref{tab_C}: i) using our fits for $A_1$  (red disks); and ii) using results found in the literature (green and blue disks).(b) Value of B obtained by: i) Evolution of $A_1/(C T^*)$ with doping, showing that it stabilizes in the value of $B$ given by Eq.~\eqref{eq_B} using our fits for $A_1$  (red disks) as well as using results found in the literature (green and blue disks); ii) the resistivity coefficient of the FL phase; and iii) the theoretical expression given in Eq.~\eqref{eq_B}. These show a remarkable agreement close to the critical point $x=\tilde{x}_0\approx 0.267$. The experimental data points were extracted from the values of the coefficient, availabe in the literature \cite{nakamae03, cooper2009anomalous, ando2004electronic, hussey2011dichotomy}. The dashed line corresponds to our theoretical calculation.}
	\label{fig_CB}
\end{figure}

	 This analysis can be consistently repeated for different families of cuprates, as shown on Fig.~\ref{fig_alpha_Ts}, and the values of $C_0$ thereby obtained are presented in Table~\ref{tab_C}. The tendency of $C$ to become a constant is visible directly in Fig.~\ref{fig_CB} (a), where we plot $A_1/(B T^*)$, with $B$ given by Eq.\eqref{eq_B}. 	

Two important results obtained in this work are the theoretical calculation of the coefficients of the  $T^2$ and $T^{1.6}$ expressions for the resistivity, respectively in the FL phase, using Eq.~\eqref{eq_B} and in the crossover regime, using Eq.~\eqref{eq_rho_cross}. 
		
		Indeed, in the FL case if we insert the value $d=6.61$\AA\  for LSCO, taken from Ref.~\cite{harshman92}, we get from Eq.~\eqref{eq_B}, $B\simeq\  0.0024\  \mu \Omega \text{cm}/K^2$. This should be compared with the reported experimental value: $B_{exp}= 0.0025\pm 0.0001 \ \mu \Omega \text{cm}/K^2$ \cite{nakamae03}. 
		
		The linear coefficient $A_1$ is directly proportional to $B$, namely, $A_1= B C T^*$, we can also compare how the values of $B$ obtained through the coefficient $A_1$, with the theoretical expression of Eq.~\eqref{eq_B}. The results of this analysis are present on Fig.~\ref{fig_CB} (b). We see that as we approach the critical point, the two definitions of $B$ coincide and are very close to our theoretical prediction. 

		The LSCO resistivity in the crossover regime, conversely, has been experimentally shown to follow the power-law $B_C T^{1.6}$. Within our theoretical approach, the crossover resistivity coefficient, $B_C$, according to (\ref{eq_rho_cross}), is then given by $B_C=B\left(C_0 T'^*(x)\right)^{0.4}$.
		
	 Using the value of $C_0$ given in Table~\ref{tab_C}, namely, $C_0=5.35$ and $T'^*(x=0.30)\simeq 42.5 K$ \cite{marino2020superconducting}, we find $B_C \simeq 0.021  \mu \Omega \text{cm}/K^{1.6}$. This should be compared with the experimental value, taken from \cite{nakamae03}, namely, $B_{C,exp} \simeq 0.019\   \mu \Omega \text{cm}/K^{1.6}$ \cite{nakamae03}.  
	 
\begin{table}[!htb]
	\centering
	\begin{tabular}{|c|c|c|}
		\hline
		Compound&$d$(\AA)& $C_0$\\
		\hline
		LSCO&6.61 \cite{harshman92}& 5.35\\
		\hline
		Bi2201&12.15 \cite{harshman92}& 5.40\\
		\hline
		Bi2212&7.74 \cite{harshman92}& 8.49\\
		\hline
		Hg1212&6.32 \cite{hunter1994pressure}& 11.41\\
		\hline
	\end{tabular}
	\caption{Values of $d$ for many compounds used to obtain the values of $C$ that were used in Fig.~\ref{fig_alpha_Ts}. The references where the values of $d$ were obtained are listed.}
	\label{tab_C}
\end{table}

{\bf Conclusions}  Starting from the model introduced in Ref.~\cite{marino2020superconducting}, and using the Kubo formula at a finite temperature, we have derived a general expression for the resistivity of High-Tc cuprates, (\ref{r}) whose particular forms in each phase namely: PG, SM, FL, as well as the crossover between the two latter,
reproduces the experimentally observed resistivity in such phases.

We calculate the resistivity coefficients in the $T$, $T^2$ and $T^{1.6}$ regimes,
our results being in agreement with the observed experimental values.
The obtainment of the  resistivity coefficients, $B$ and $B_C$, in particular, without adjusting any parameter, attests the accuracy of our model for the description of High-Tc cuprates.

Based on this model, we may conclude that the main cause of resistivity in cuprates is the scattering of the charged holes by excitons, which are associated with a scalar field. The presence of excitons should be expected in a system containing electrons and holes. Being confined to the $CuO_2$ planes, they should present similar properties as, for instance, the exciton states observed in transition metal dichalcogenides by photoluminescence techniques, namely, high binding energy and very short lifetimes  \cite{TMDexcitons}.

Our expression for the resistivity in cuprates involves a two-variable scaling function. The SM phase, where the linear resistivity is seen, appears as a quantum critical region associated to a quantum critical point located at the right extremity of the SC dome, precisely where the $T_c$, $T^*$, $T'^*$ and $T_{\text{cross}}$ lines meet. Quantum criticality results from the loss of the energy scale $M$, related to the exciton scattering as $M$ goes to zero when we approach the transition line $T^*(x)$ separating the PG from the SM phase. The scaling of every energy with $T$ in the quantum critical phase, makes the resistivity to become linear.

  The possibility of expressing the resistivity in the SM phase in terms of a scaling function was already pointed out in the literature \cite{phillips05, sachdev97, phillips2012advanced, sachdev_2011}. In this work, however, this fact has been derived from the proposed Hamiltonian, and the explicit form of the scaling function was, thereby, determined. Also, the resistivity being the consequence of hole scattering by a scalar field, our scaling approach does not suffer from the problems pointed out in \cite{phillips05, phillips2012advanced}.	We, therefore, reconcile the existence of a quantum critical regime with the Planckian diffusion hypothesis.
  
  The scaling allows us to predict that the slope of the linear resistivity is proportional to the PG temperature $T^*(x)$, where the excitons condense. We, thereby, can infer the connection between the resistivity and the scattering by excitons.


	A natural extension of this work would be to consider the inclusion of a third scaling variable, representing the effect of external agents such as pressure or magnetic field on the resistivity of cuprates. That would add a third dimension to the phase diagram of Fig.~\ref{fig_phase_diagram}. It would be interesting to compare the new results with the data available in the literature for external magnetic field and with an AC field. The effect of pressure, that changes $\mu$ ~\cite{marino2020superconducting}, on resistivity is also an interesting direction to be explored using this formalism.

{\bf Acknowledgments} RA acknowledges funding from CAPES. ECM was supported in part by CNPq and FAPERJ.

\bibliography{cuprates}

\clearpage
\newpage

\begin{widetext}
\begin{center}
\textbf{\large Supplemental Material}
\end{center}
\setcounter{equation}{0}
\setcounter{figure}{0}
\setcounter{table}{0}
\setcounter{section}{0}
\setcounter{page}{1}
\renewcommand{\theequation}{S\arabic{equation}}
\renewcommand{\thefigure}{S\arabic{figure}}
\renewcommand{\thetable}{S\Roman{table}}
\renewcommand{\thesection}{S\Roman{section}}
\section{Brief Review of the Model}
We use here the same Hamiltonian introduced in \cite{marino2020superconducting} , namely
	\begin{eqnarray}
			&&H_{eff}[\psi]=-t \sum_{\textbf{R},\textbf{d}_i} \psi_{A\sigma}^\dagger\left(\textbf{R}\right)\psi_{B\sigma}\left(\textbf{R}+\textbf{d}_i\right)+hc\nonumber\\
			&  &-g_S\sum_{\textbf{R},\textbf{d}_i} \left[\psi_{A\uparrow}^\dagger\left(\textbf{R}\right)\psi_{B\downarrow}^\dagger\left(\textbf{R}+\textbf{d}_i\right)+ \psi^\dagger_{B\uparrow}\left(\textbf{R}+\textbf{d}_i\right)\psi_{A\downarrow}^\dagger\left(\textbf{R}\right)\right]\left[\psi_{B\downarrow}^{\ }\left(\textbf{R}+\textbf{d}_i\right)\psi_{A\uparrow}\left(\textbf{R}\right)+ \psi_{A\downarrow}\left(\textbf{R}\right) \psi_{B\uparrow}\left(\textbf{R}+\textbf{d}_i\right)\right]\nonumber \\
			&  &-g_P\sum_{\textbf{R},\textbf{d}_i} \left[\psi_{A\uparrow}^\dagger\left(\textbf{R}\right)\psi_{B\uparrow}\left(\textbf{R}+\textbf{d}_i\right)+\psi_{A\downarrow}^\dagger\left(\textbf{R}\right) \psi_{B\downarrow}\left(\textbf{R}+\textbf{d}_i\right)\right]\left[\psi_{B\uparrow}^\dagger\left(\textbf{R}+\textbf{d}_i\right)\psi_{A\uparrow}\left(\textbf{R}\right)+\psi_{B\downarrow}^\dagger\left(\textbf{R}+\textbf{d}_i\right) \psi_{A\downarrow}\left(\textbf{R}\right)\right],
		\label{eq_ham_SF_Hubbard}
	\end{eqnarray}
in the above expression, $\textbf{R}$ denotes the sites of a square sublattice (A) and $\textbf{d}_i$, $i=1,...,4$, its nearest neighbors, belonging to square sublattice (B). $ \psi_{A,B\sigma}^\dagger$ is the creation operator of a hole, or, equivalently, the destruction operator of an electron, with spin $\sigma=\uparrow,\downarrow$ in sublattice $A,B$. Such sublattices are formed as follows: each oxygen ion possesses a $p_x$ and a $p_y$ orbital but only one of them, either $p_x$ or $p_y$, alternatively, hybridizes with the copper 3d orbitals . Two inequivalent oxygen sublattices are thereby formed, one having hybridized $p_x$ orbitals and the other having $p_y$.
 $t$ is the usual hopping parameter, $g_S$ is the coupling parameter of the hole-attractive interaction term and $g_P$,  coupling parameter of the hole-repulsive interaction term
 . 
Through a Hubbard Stratonovitch (HS) transformation we can express this Hamiltonian in terms of the scalar HS fields $\Phi(\textbf{k})$ and $\chi(\textbf{k})$, which upon integration, generate the quartic attractive and repulsive terms, respectively responsible for the formation of Cooper pair and exciton bound states.
	\begin{eqnarray}
		&H_{eff}[\psi,\Phi,\chi]=&\sum_{\textbf{k},\sigma}\left\{\epsilon(\textbf{k}) \psi_{A\sigma}^\dagger(\textbf{k})\psi_{B\sigma}(\textbf{k})+ \Phi(\textbf{k}) \Big [\psi_{A\uparrow}^\dagger(-\textbf{k})\psi^\dagger_{B\downarrow}(\textbf{k}) +\psi^\dagger_{B\uparrow}(\textbf{k}) \psi_{A\downarrow}^\dagger(-\textbf{k})\Big]+\chi(\textbf{k}) \Big [\psi_{A\sigma}^\dagger(\textbf{k})\psi_{B\sigma}(\textbf{k})  \Big ]+ hc\right\}\nonumber \\
		&&+ \frac{1}{g_S}\sum_{\textbf{k}}\Phi^\dagger(\textbf{k}) \Phi(\textbf{k})+ \frac{1}{g_P}\sum_{\textbf{k}}\chi^\dagger(\textbf{k}) \chi(\textbf{k}).
		\label{eq_ham_SF_Hubbard_HS}
	\end{eqnarray}
The ground-state expectation values of the HS fields, namely, $\Delta=\langle \Phi\rangle$ and $M=\langle \chi\rangle$ are the order parameters for the superconducting (SC) and pseudogap (PG) phases \cite{marino2020superconducting}. 

We introduce the doping $x$-dependence through the constraint
\begin{eqnarray}
\lambda\Big[ \sum_{C=A,B} \psi^\dagger_{C,\sigma,a} \psi_{C,\sigma,a} - N d(x)\Big ]
\label{5a}
\end{eqnarray}
which is implemented by integrating over the Lagrange multiplier field $\lambda$, whose vacuum expectation value is the chemical potential:
 $\langle \lambda \rangle = \mu$.
Here $d\left(x\right)$ is a function of the stoichiometric doping parameter, which turns out to be $d\left(x\right)=2\left(x_0-x\right)/x_0$ and $N$ is the number of $CuO_2$ planes intercepting the material primitive unit cell. \cite{marino2020superconducting}.

	We can represent Hamiltonian~\eqref{eq_ham_SF_Hubbard_HS} with the constraint of Eq.~\eqref{5a}, using Nambu fermion fields $\Psi_{a}$
	\begin{equation}
		\Psi_{a}=\left(\begin{matrix}
				\psi_{A,\uparrow,a} \\
				\psi_{B,\uparrow,a}  \\
				\psi^\dagger_{A,\downarrow,a} \\
				\psi^\dagger_{B,\downarrow,a}   
		\end{matrix}\right),
		\label{eq_nambu}
	\end{equation}
as
	\begin{equation}
		H_{eff}-\mu \mathcal{N} =\frac{1}{g_S}\sum_{\textbf{k}}|\Delta(\textbf{k})|^2 +  \frac{1}{g_P}\sum_{\textbf{k}}|M(\textbf{k})|^2+ \sum_{\textbf{k}} \Psi^{\dagger}_{a}(k) \left(\mathcal{H}\left(k\right) -\mu \mathcal{N}\right)\Psi_{a}(k)
		\label{eq_ham_nambu}
	\end{equation}
with the matrix $\mathcal{H}-\mu \mathcal{N}$
	\begin{eqnarray}
		\mathcal{H}-\mu \mathcal{N}=\left(\begin{matrix}
			-\mu & \epsilon + M & 0 &  \Delta\\
			\epsilon + M^* & -\mu & \Delta & 0\\
			0 & \Delta^{*} & \mu & -\epsilon -M^*\\
			\Delta^{*} & 0 & -\epsilon -M & \mu\\
		\end{matrix}\right)\, .
	\label{M2}
	\end{eqnarray}

By integrating on the fermion fields, we obtain the grand-partition functional
as well as the grand-canonical potential $\Omega [ \Delta,M,\mu  ]$:
$$
Z=\exp\Big \{ -\beta \Omega [ \Delta,M,\mu  ]\Big \}
$$ 

Minimizing the grand-canonical potential $\Omega$, respectively, with respect to $\Delta,M,\mu $, we obtain the equations that determine the behavior of $\Delta$, $M$ and $\mu$ in the thermodynamic equilibrium state \cite{marino2020superconducting}:
\begin{eqnarray}
&2\Delta\left[-\frac{2T}{\alpha}F\left(\Delta,M,\mu \right)+\frac{\eta\left(N g_S\right)}{g_c}\right]=0\label{eq_Delta}\\
&2M\left[-\frac{2T}{\alpha}F\left(\Delta,M,\mu \right)+\frac{\eta\left(N g_P\right)}{g_c}\right]=0\label{eq_M}\\
&d\left(x\right)=\mu \frac{4T}{\alpha}F\left(\Delta,M,\mu \right), 
\label{eq_dx_mu}
\end{eqnarray}
with the critical coupling $g_c=0.30$ eV and $F$ being, close to the critical curves, given by
	\begin{equation}
		\left.F\left(\Delta,M,\mu \right)\right|_{\left|\Delta\right|\sim 0, \left|M\right|\sim 0}=\ln 2+\frac{1}{2}\ln\cosh\left[\frac{\sqrt{\left|\Delta\right|^2+\left(\left|M\right|+\mu\left(x\right)\right)^2}}{2T}\right]+\frac{1}{2}\ln\cosh\left[\frac{\sqrt{\left|\Delta\right|^2+\left(\left|M\right|-\mu\left(x\right)\right)^2}}{2T}\right].
	\end{equation}

	In the superconducting phase, $\Delta \neq 0$ and $M=0$ so that Eq.~\eqref{eq_M} is trivially satisfied and from Eqs.~\eqref{eq_Delta} and ~\eqref{eq_dx_mu} we arrive at the expressions for the superconducting transition temperature $T_c\left(x\right)$:
	\begin{equation}
		\begin{cases}
			T_{c}(x) =\frac{\ln2 \,\ T_{max}}{\ln2 + \frac{\mu_0(x)}{2T_{c}(x)} + \frac{1}{2}\left(e^{-\frac{\mu_0(x)}{T_{c}(x)}}-1\right)},\hspace{0.5cm}x < x_{0}\\
			T_c(x) =\frac{\ln2 \ \ T_{max}}{\ln\Big [1+ \exp\left[- \frac{\mu_0(x)}{T_c(x)} \right]  \Big ]},\hspace{1.6cm} x > x_{0}\\
			T_{c}(x) =\frac{\ln2 \,\ T_{max}}{\ln2 + \frac{|\mu_0(x)|}{2T_{c}(x)} + \frac{1}{2}\left(e^{-\frac{|\mu_0(x)|}{T_{c}(x)}}-1\right)},\hspace{0.3cm} \text{LSCO},
			\label{eq_Tc}
		\end{cases}
	\end{equation}
where
\begin{equation}
\mu_0\left(x\right)=2\gamma\left(x_0-x\right)
			\label{mu}
	\end{equation}
  $\gamma$ being a parameter to be determined for each compound. $T_{\max}=\frac{\Lambda}{2\ln 2}\eta\left(N g_S\right)$, $\Lambda=0.018$ eV is an energy cut-off and $\eta\left(N g_S\right)=1-\frac{g_c}{N g_S}$. Notice that for LSCO we use a symmetrized version of the equations to comply with the experimental observation of a symmetrical SC dome.

	For the pseudogap phase, $\Delta=0$ and $M\neq 0$ so that now Eq.~\eqref{eq_Delta} is trivially satisfied and Eqs.~\eqref{eq_M} and ~\eqref{eq_dx_mu} imply for the pseudogap transition temperature $T^*\left(x\right)$: 
	\begin{equation}
		T^*(x) =\frac{\frac{\Lambda\eta(g_PN)}{2}}{\ln\Big [1+ \exp\left[- \frac{2\tilde\gamma (\tilde x_0 - x)}{T^*(x)} \right]  \Big ]},
		\label{eq_Ts}
	\end{equation}
with $\eta\left(g_PN\right)=1-\frac{g_c}{N g_P}$, $\tilde x_0=x_{SC}^{+}$ (see Main Text) and $\tilde\gamma$ is determined for each compound.

	Using a systematic procedure, discussed extensively in Ref.~\cite{marino2020superconducting}, we were able to obtain the parameters $\gamma$, $g_S$, $g_P$ and $\tilde{\gamma}$ for each material and characterize their phase diagram with an excellent agreement with experimental data.

which yields the grand-canonical potential in the presence of an applied electromagnetic vector potential $ \textbf{A}$, namely, $\Omega [ \textbf{A} ]$.

\section{Complete Resistivity Calculation}

The DC conductivity, at a finite temperature, according to the Kubo formula \cite{mahan2013many}, is given by
\begin{eqnarray}
\sigma^{ij}_{\text{DC}}=\lim\limits_{\omega\rightarrow 0}\frac{1}{ \omega}\left[1- e^{-\beta\hbar\omega} \right ]\lim\limits_{\mathbf{k}\rightarrow \mathbf{0}}\langle j^{i}j^{j}\rangle_{\text{C}}\left(\omega, \mathbf{k}\right).
\label{10z}
\end{eqnarray}

The average electric current and its two-point correlator are obtained from:
\begin{eqnarray}
\langle j^i \rangle = \frac{\delta  \Omega [ \textbf{A} ] }{\delta \textbf{A}^i}\ \ \ \ ,\langle j^i j^j\rangle = \frac{\delta^2  \Omega [ \textbf{A} ] }{\delta \textbf{A}^i \delta \textbf{A}^j}.
\label{eq_j}
\end{eqnarray}
where $\Omega [ \textbf{A} ]$ is the grand-canonical potential in the presence of an applied electromagnetic vector potential $ \textbf{A}$. This relates to the grand-partition functional $Z[\textbf{A}]$ as
\begin{eqnarray}
\Omega [ \textbf{A} ]=-\frac{1}{\beta} \ln Z[\textbf{A}] .
 \label{2}
\end{eqnarray}
The grand-partition functional, is given by
\begin{eqnarray}
Z[\textbf{A}]= {\rm Tr} e^{-\beta \left[ H[\textbf{A}]-\mu\mathcal{N}\right ]}
 \label{3}
\end{eqnarray}
where $\mu$ is the chemical potential, $\mathcal{N}$ is the number operator and the electromagnetic field $\textbf{A}$ is introduced through the usual minimal coupling prescription
\begin{eqnarray}
\epsilon(\textbf{k}) \longrightarrow \epsilon(\textbf{k} +
 e \textbf{A})
\label{4}
\end{eqnarray}
where $\epsilon(\textbf{k})=2t [ \cos k_x a + \cos k_y a ]$ is the usual tight-binding energy for a square lattice.

We can write the eigenvalues of $H-\mu\mathcal N$, in terms of the stationary values $\Delta_0; M_0; \mu$ as
\begin{eqnarray}
 \mathcal{E}_{\pm}(\textbf{k}))=\pm \sqrt{\left(\sqrt{\epsilon^2(\textbf{k})+ M_0^2}\pm \mu\right)^2 +\Delta_0^2}.
\label{7}
\end{eqnarray}

The grand-partition functional $Z[\textbf{A}]$ follows from (\ref{3}) and Eq.\eqref{4}, namely,
\begin{eqnarray}
Z[\textbf{A}]&=&\exp\left\{-\beta\left\{ \frac{\left|\Delta\right|^2}{g_S}+\frac{\left|M\right|^2}{g_P}+N\mu\left(x\right)-NTA\sum\limits_{n=-\infty}^{\infty}\sum\limits_{l=\pm 1}\int \frac{d^2k}{4\pi^2}\ln\left[\omega_n^2+\mathcal{E}_{l}^2[\textbf A]\right]\right\}\right\}\\
	&=&\exp \left \{ -\beta T\sum_{\omega_n}\sum_{l=\pm 1}\int \frac{d^2k}{(2\pi)^2} \ln\left[\frac{ \omega_n^2 +  \mathcal{E}_{l}^2[\textbf A]}{ \omega_n^2 +  \mathcal{E}_{l}^2[0]} \right ]\right \} 
 \nonumber \\
\label{8}
\end{eqnarray}
where
\begin{eqnarray}
  \mathcal{E}_{l}^2[\textbf A]=\Delta_0^2+\Big (\sqrt{v^2 (\textbf k +e \textbf A)^2 + M_0^2} + l\mu \Big )^2.
\label{9}
\end{eqnarray}
Using (\ref{eq_j}), (\ref{4}) and (\ref{9}), we obtain the average current:
\begin{eqnarray}
\langle j^i \rangle \left(\textbf{k}=0, \omega=0\right)= N  \sum_{l =\pm 1} 2T  \mathcal{E}_{l}[\textbf A] \frac{\partial   \mathcal{E}_{l}[\textbf A]}{\partial \textbf {A}^i}
\sum_{\omega_n}\frac{1}{ \omega_n^2 +   \mathcal{E}_{l}^2[\textbf A]}=N \sum_{l =\pm 1} \frac{\partial   \mathcal{E}_{l}[\textbf A]}{\partial \textbf {A}^i} \tanh\Big (\frac{  \mathcal{E}_{l}[\textbf A]}{2k_BT} \Big ).
\label{10}
\end{eqnarray}
In order to obtain the conductivity matrix, $\sigma^{ij}$ we must take the derivative of $\langle j^i \rangle$ with respect to $\textbf {A}^j$, at $\textbf {A}= \textbf {k}=0$. Considering that, in this case 
$$
\left.\frac{\partial   \mathcal{E}[\textbf A]}{\partial \textbf {A}^i}\right|_{\textbf{A}=0, \textbf{k}=0}=0
$$
we have
\begin{eqnarray}
\langle j^i j^j\rangle\left(\textbf{k}=0, \omega=0\right) =
 N \sum_{l =\pm 1} \frac{\partial^2   \mathcal{E}_{l}[\textbf A]}{\partial \textbf {A}^i \partial \textbf {A}^j} \tanh\Big (\frac{  \mathcal{E}_{l}[\textbf A]}{2k_BT} \Big ).
\label{10y}
\end{eqnarray}
Under the latter conditions, only the diagonal ($\delta^{ij}$) terms survive, namely
\begin{eqnarray}
&&\hspace{-0.5cm}\langle j^i j^j \rangle\left(\textbf{k}=0, \omega=0\right) = \frac{N e^2 v^2 }{M_0} \delta^{ij}\left \{ \frac{(M_0+\mu)\tanh\left[\frac{\sqrt{\Delta_0^2+\Big ( M_0 + \mu \Big )^2}}{2k_BT}  \right]}{\sqrt{\Delta_0^2+\Big ( M_0 + \mu \Big )^2}}+ \right .\left .  \frac{(M_0-\mu)\tanh\left[\frac{\sqrt{\Delta_0^2+\Big ( M_0 - \mu \Big )^2}}{k_BT}  \right]}{\sqrt{\Delta_0^2+\Big ( M_0 - \mu \Big )^2}} \right \}
\label{11}
\end{eqnarray}

 In order to obtain the DC conductivity per $CuO_2$ plane, we just divide by $N$.
The corresponding DC resistivity per $CuO_2$ plane, then, will be given by
\begin{eqnarray}
&\ &\rho^{ij} =\left(\frac{\sigma^{ij}_{\text{DC}}}{N}\right)^{-1}\hspace{-0.5cm}=\delta^{ij}\frac{M_0}{\hbar\beta V^{-1} e^2 v^2 } \left \{ \frac{(M_0+\mu)\tanh\left [\frac{\sqrt{\Delta_0^2+\Big ( M_0 + \mu \Big )^2}}{2k_BT}  \right ]}{\sqrt{\Delta_0^2+\Big ( M_0 + \mu \Big )^2}}+ \right .\left .  \frac{(M_0-\mu)\tanh\left [\frac{\sqrt{\Delta_0^2+\Big ( M_0 - \mu \Big )^2}}{2k_BT}  \right ]}{\sqrt{\Delta_0^2+\Big ( M_0 - \mu \Big )^2}} \right \}^{-1}
\label{11y}
\end{eqnarray}
where $V=da^2$ is the primitive unit cell volume per  $CuO_2$ plane, that is introduced through the Fourier transform.

Then, considering that ${\rm sign}(x)\tanh|x|=\tanh (x)$, we find that the resistivity for $\Delta=0$ is given by
\begin{eqnarray}
\rho^{ij} = \delta^{ij} \frac{V}{\hbar \beta e^2v^2}
 \left [\frac{M_0 }{\tanh\left (\frac{ M_0 + \mu }{2k_BT}  \right ) +
 \tanh\left (\frac{ M_0 - \mu }{2k_BT}  \right )} \right ].
\label{12x}
\end{eqnarray}
This can be rewritten as
\begin{eqnarray}
\rho^{ij} &=& \delta^{ij} \frac{Vk_B}{\hbar v^2e^2}T \frac{M_0\left[\cosh\left(\frac{M_0}{k_BT}\right)+\cosh\left(\frac{\mu}{k_BT}\right)\right]}{2\sinh\left(\frac{M_0}{k_BT}\right)} ,
\nonumber \\
\label{eq_res_SM}
\end{eqnarray}

that is precisely Eq.~\eqref{eq_rho} of the main text.

\subsection{Particular Forms of $G(K_1,K_2)$ in All Phases}

As discussed in the main text, the different behavior of the resistivity, observed in each phase can be understood as different particular cases of the scaling function $G(K_1,K_2)$
		\begin{equation}
		G\left(\frac{M}{k_B T}, \frac{\mu}{k_B T}\right)=\frac{M}{k_B T}\frac{\cosh\left(\frac{M}{k_B T}\right) +\cosh\left(\frac{\mu}{k_B T}\right)}{2\sinh \left(\frac{M}{k_B T}\right)}
	\end{equation}
in the different phases. Here we derive the particular expressions of $G$ in the different regions.

	\begin{description}
		\item[Pseudogap Phase, $T\rightarrow 0$]{
		
		As $M$ and $\mu$ are finite for zero temperature in this phase we can approximate
		\begin{eqnarray}
			\frac{\cosh\left(\frac{M}{k_B T}\right) +\cosh\left(\frac{\mu}{k_B T}\right)}{\sinh \left(\frac{M}{k_B T}\right)}&=&\frac{\exp\left(\frac{M}{k_B T}\right)+\exp\left(-\frac{M}{k_B T}\right)+\exp\left(\frac{\mu}{k_B T}\right)+\exp\left(-\frac{\mu}{k_B T}\right)}{\exp\left(\frac{M}{k_B T}\right)-\exp\left(-\frac{M}{k_B T}\right)}\nonumber\\ 
			&=&\frac{1+\underbrace{\exp\left(-\frac{2 M}{k_B T}\right)}_{\approx 0}+\exp\left(\frac{\mu-M}{k_B T}\right)+\underbrace{\exp\left(-\frac{\mu+ M}{k_B T}\right)}_{\approx 0}}{1-\underbrace{\exp\left(-\frac{2 M}{k_B T}\right)}_{\approx 0}}\nonumber\\ 
			&\approx&1+\exp\left(\frac{\mu-M}{k_B T}\right)\nonumber\\	
			&\approx&\exp\left(\frac{\mu-M}{k_B T}\right),
		\end{eqnarray}
such that 
		\begin{equation}
			G_{PG}( T\rightarrow 0)\approx\frac{M}{2 k_B T} \exp\left(\frac{\mu-M}{k_B T}\right)\equiv \frac{M}{2 k_B T} \exp\left(\frac{T_0}{T}\right),
		\end{equation}
where we define the temperature scale $T_0\equiv \frac{\mu-M}{k_B}$.
		}
		\item[Pseudogap Phase, $T\rightarrow T^*$]{
		
		For $T\rightarrow T^*$, $M\rightarrow 0$. We can separate inspect the limit of two separate parts of $F$ in this limit. First, the sum of $\cosh$,
		\begin{equation}
			\lim\limits_{T\rightarrow T^*}\frac{\cosh\left(\frac{M}{k_B T}\right) +\cosh\left(\frac{\mu}{k_B T}\right)}{2}=\frac{1+\cosh\left(\frac{\mu_{PG} (T\rightarrow T^*)}{k_B T^*}\right)}{2}=\cosh^2\left(\frac{\mu_{PG} (T\rightarrow T^*)}{2 k_B T^*}\right),
		\end{equation}	
then the fraction
		\begin{equation}
			\lim\limits_{T\rightarrow T^*}\frac{M/k_B}{\sin\left(\frac{M}{k_B T}\right)}=T^*,
		\end{equation}
such that
		\begin{equation}
			G_{PG} (T\rightarrow T^*)=\frac{T^*}{T}\cosh^2\left(\frac{\mu_{PG} (T\rightarrow T^*)}{2 k_B T^*}\right).
		\end{equation}
		}
		\item[Strange Metal Phase]{
		
		In this phase, $M\rightarrow 0$ for any temperature. This imply that $F$ will have the same behavior of the previous phase, but with a different dependence of the $\cosh$ 
		\begin{equation}
			G_{SM}=C_0\frac{T^*}{T},
		\end{equation}
where $T^*$ comes from the condition of the continuity of $\rho$ at $T=T^*$ and $C_0$ is the value of $\cosh^2\left(\frac{\mu}{2 k_B T}\right)$ in the strange metal phase as discussed in the main text. 
		}
		\item[Fermi Liquid Phase]
		
		{	
		For the Fermi Liquid phase, the limit $M\rightarrow 0$ is taken without the constraint imposed by the continuity of $\rho$ at $T^*$. 
			On the other hand, for $\Delta=M=0$, equations (\ref{eq_Delta}) and (\ref{eq_M}) are trivially satisfied, and Eq. (\ref{eq_dx_mu}) implies
\begin{equation}
			\frac{\mu_{FL}}{2T}\left[\mu^2_{FL}-\Lambda\tilde{\eta} \tilde{\mu}(x)\right] = 0,
			\label{eqmu}
		\end{equation}
where  $\tilde{\mu}(x)= 2\tilde{\gamma}(\tilde{x}_0-x)$.

Observe that, consequently, for $x\rightarrow \tilde{x}_0$, we have the solution of the above equation:  $\mu_{FL}(x) = 0$. 

Now, for $x >\tilde{x}_0$, we see that $\tilde{\mu}(x)$ becomes negative, hence the only solution of (\ref{eqmu}) is
$\mu_{FL}(x) = 0$ as well. 

Hence, in the FL phase, we have
		\begin{equation}
			G_{FL}=1.
		\end{equation}
		}
	\end{description}


\section{Data Analysis}
\subsection{Determination of the Doping Level}

For compounds other than LSCO, the actual in-plane doping level is not directly related to the stoichiometric
doping, due to the absence of a one-to-one relation between the amount of doped atomic species and the number of
holes actually introduced into the planes. This creates a problem, since the chemical potential depends on the latter
while measured physical quantities, are expressed in terms on the former. In Ref.~\cite{marino2020superconducting} a solution for this problem
was obtained. Noting that the two quantities, although usually different, are closely related, it was assumed that
the chemical potential should be proportional to the stoichiometric doping, as we can see in (\ref{mu}). The proportionality coefficient, $\gamma$ was then
determined for each compound by comparing the results thereby obtained with the experimental data. That is how
the curves $T_c(x)$ and $T^*(x)$ were obtained, with excellent agreement to the experiments for several cuprate compounds.
Nevertheless, for some of the experimental data we have analyzed, a uniform database of doping parameters for the
whole set of samples was needed. This is so because the reported doping levels were either based on different analyses
or simply were not provided for each sample, but only its critical temperature. To circumvent this problem, we used
our theoretical expression for $T_c(x)$, derived in Ref.~\cite{marino2020superconducting} in order to obtain the doping level of each sample by using
the corresponding value of $T_c$. The resulting $x$ values are presented on Table.~\ref{tab_calib}.


\begin{table}[!htb]
	\centering
	\begin{tabular}{|c|c|c|c|c|c|c|c|c|c|c|c|c|c|}
		\hline
		Compound&$T_c$& $x$& Ref.&\hspace{0.5cm}&Compound&$T_c$& $x$& Ref.&\hspace{0.5cm}&Compound&$T_c$& $x$& Ref.\\
		\hline
		Bi2201&$14.5$ & $ 0.14$ & \cite{ando2000carrier}&&Bi2212&$35$ & $ 0.13$ & \cite{akoshima98}&&Hg1212&$86$ & $ 0.12$ & \cite{yamamotopress}\\
		Bi2201&$23.7$ & $ 0.17$ & \cite{ando2000carrier}&&Bi2212&$50$ & $ 0.14$ & \cite{akoshima98}&&Hg1212&$104$ & $ 0.15$ & \cite{yamamotopress}\\
		Bi2201&$30.2$ & $ 0.21$ & \cite{ando2000carrier}&&Bi2212&$65$ & $ 0.16$ & \cite{akoshima98}&&Hg1212&$124$ & $ 0.22$ & \cite{yamamotopress}\\
		Bi2201&$34.2$ & $ 0.26$ & \cite{ando2000carrier}&&Bi2212&$80$ & $ 0.18$ & \cite{akoshima98}&&Hg1212&$120$ & $ 0.25$ & \cite{yamamotopress}\\
		Bi2201&$33.1$ & $ 0.30$ & \cite{ando2000carrier}&&Bi2212& $87$ & $ 0.20$ & \cite{akoshima98}& &Hg1212& $89$ & $0.28$ & \cite{yamamotopress} \\
		Bi2201&$30.1$ & $ 0.31$ & \cite{ando2000carrier}&&Bi2212& $86$ & $ 0.25$ & \cite{akoshima98}& & &  && \\
		&&  & &&Bi2212 &$84$ & $0.26$ & \cite{akoshima98}&&&&& \\
		\hline
	\end{tabular}
	\caption{Obtained doping levels using Eq.\eqref{eq_Tc} for many analyzed compounds. The references where the experimental data were obtained are also listed.}
	\label{tab_calib}
\end{table}

\subsection{Fitting procedure}

	To fit the data, we have added a constant value $\rho_0$ to the theoretical expression of Eq.~\eqref{eq_rho_PG_0} and Eq.~\eqref{eq_rho_SM}, in order to account for any effect of material imperfections in the resistivity. So, in summary, we have fitted $A_1$ and $\rho_0$ in the strange metal phase and $T_0$ (assuming that it does not dependent on $T$) in the low-T regime of the PG phase. 
	
	The fitted data for LSCO, Bi2201, Bi2212 and Hg1212 are in Figs.~\ref{fig_res_LSCO},~\ref{fig_res_Bi2201}, ~\ref{fig_res_Bi2212} and ~\ref{fig_res_Hg1212}. Observe how
these expressions describe the resistivity for huge range of experimental data. The input values of the latter were
obtained by digitalization of the published data.
	\begin{figure}[!h]
		\includegraphics[width=\linewidth]{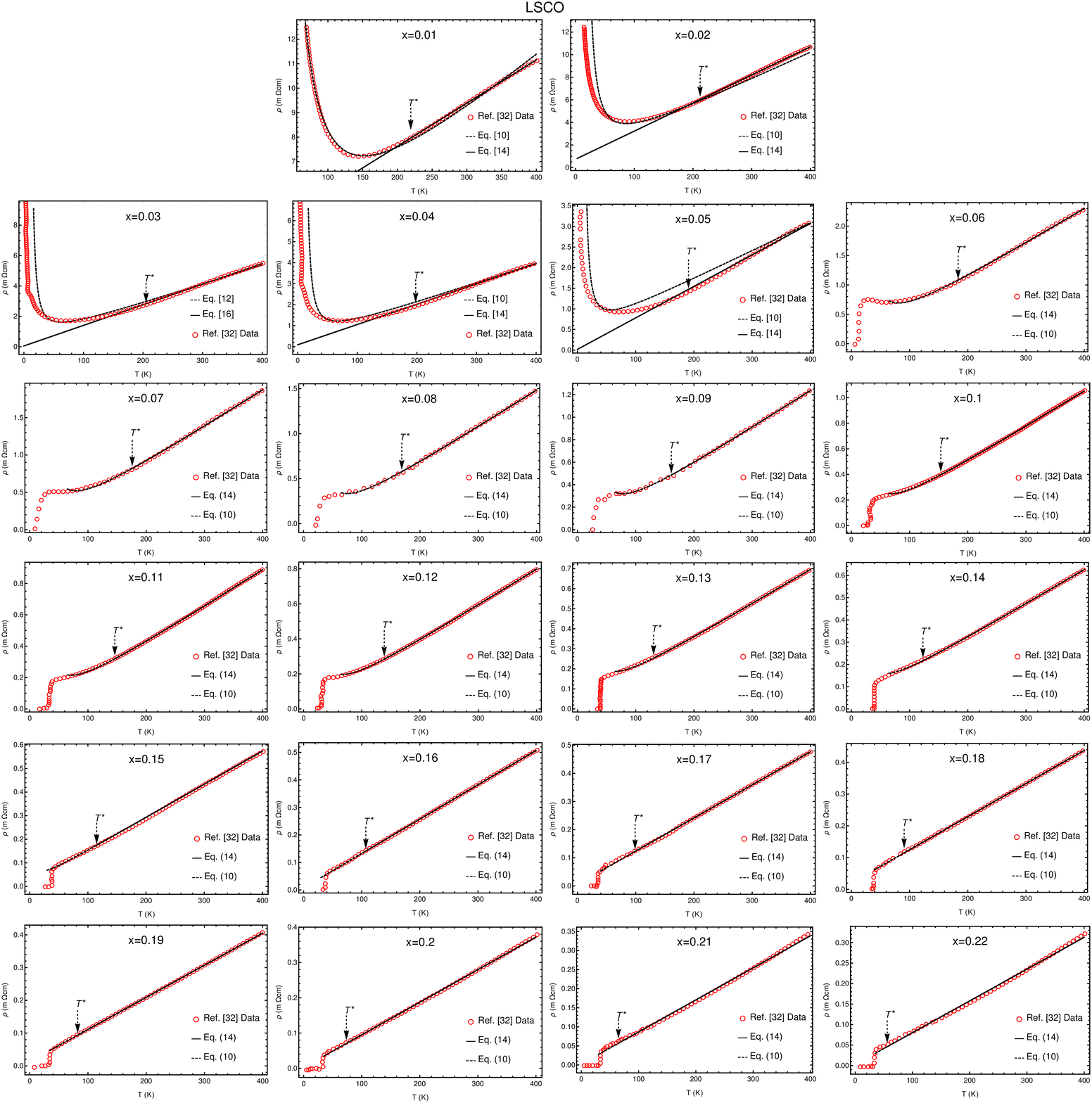}
		\caption{Comparison between the experimental data extracted from Ref.~\cite{ando2004electronic} and the fitted Eq.~\eqref{eq_rho_PG_0} and Eq.~\eqref{eq_rho_SM}(with addition of $\rho_0$) for LSCO.}
		\label{fig_res_LSCO}
	\end{figure}
	\begin{figure}[!h]
		\includegraphics[width=\linewidth]{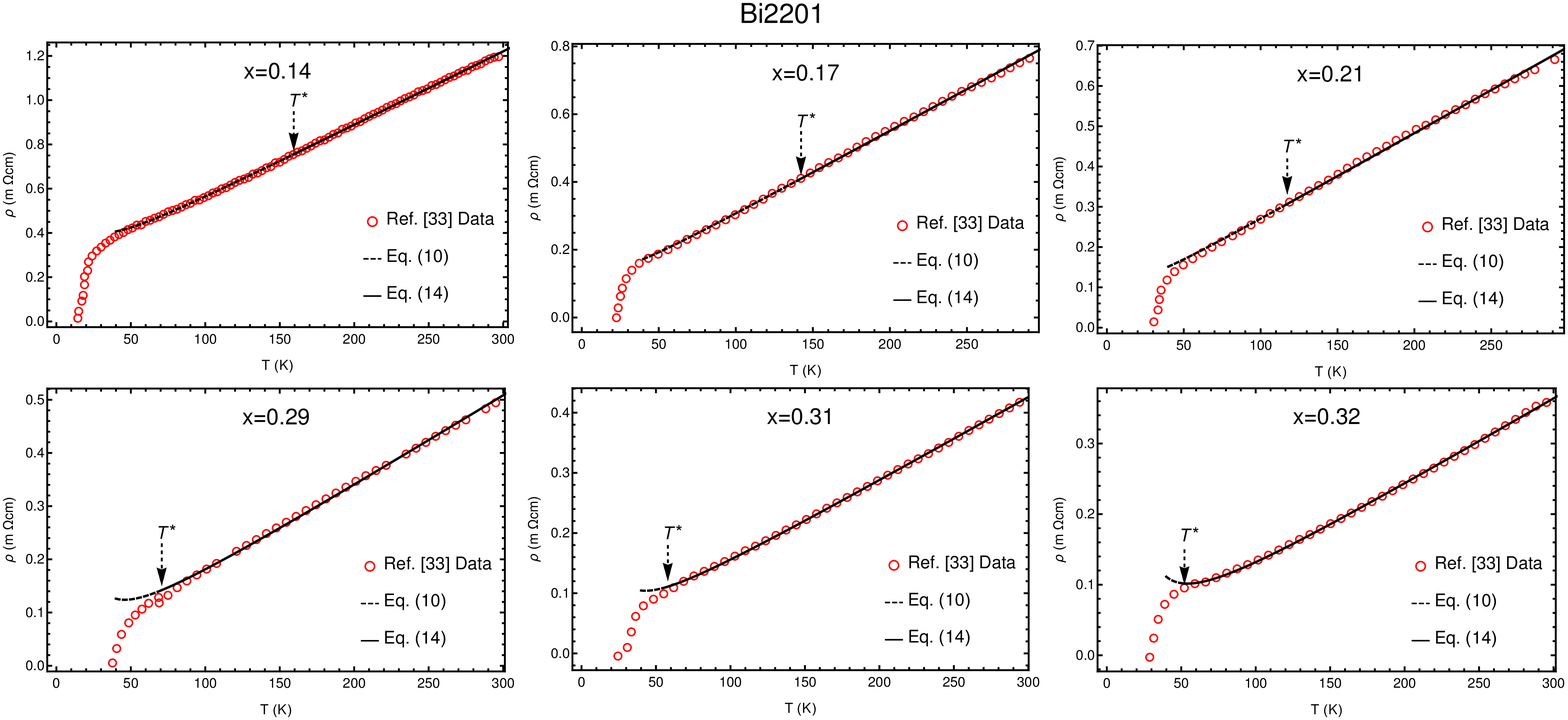}
		\caption{Comparison between the experimental data extracted from Ref.~\cite{ando2000carrier} and the fitted Eq.~\eqref{eq_rho_SM} (with addition of $\rho_0$) for Bi2201.}
		\label{fig_res_Bi2201}
	\end{figure}
	\begin{figure}[!h]
		\includegraphics[width=\linewidth]{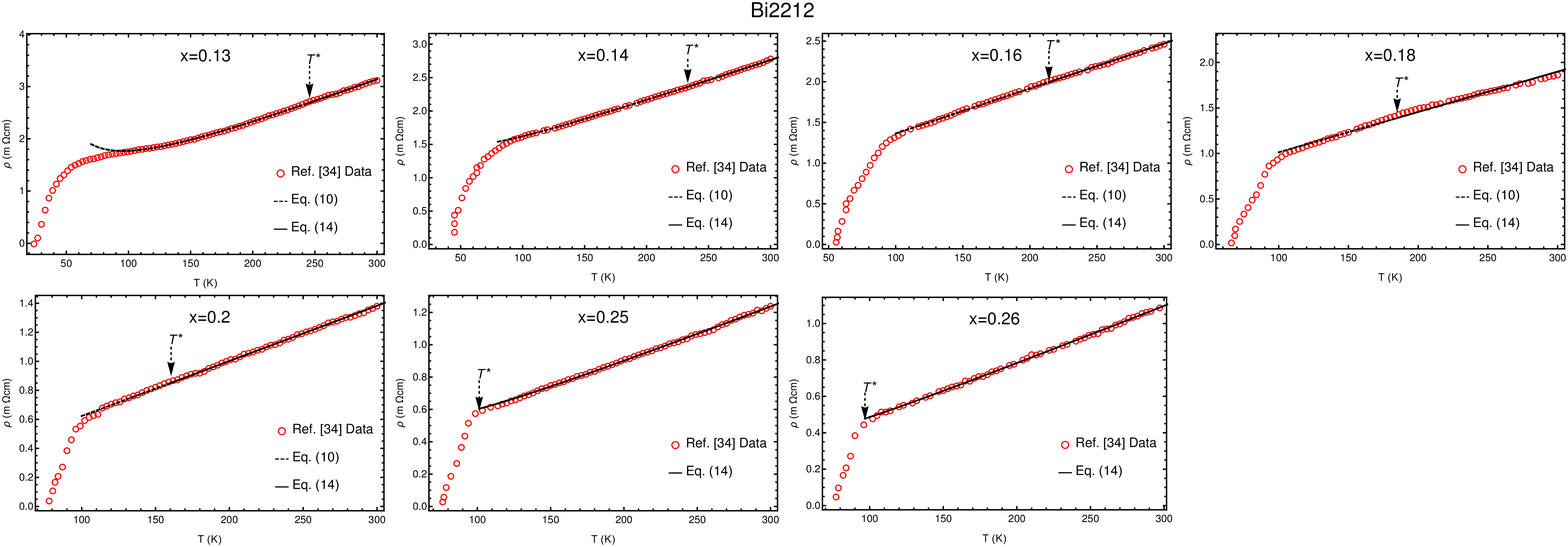}
		\caption{Comparison between the experimental data extracted from Ref.~\cite{akoshima98} and the fitted Eq.~\eqref{eq_rho_SM} (with addition of $\rho_0$) for Bi2212.}
		\label{fig_res_Bi2212}
	\end{figure}
	\begin{figure}[!h]
		\includegraphics[width=\linewidth]{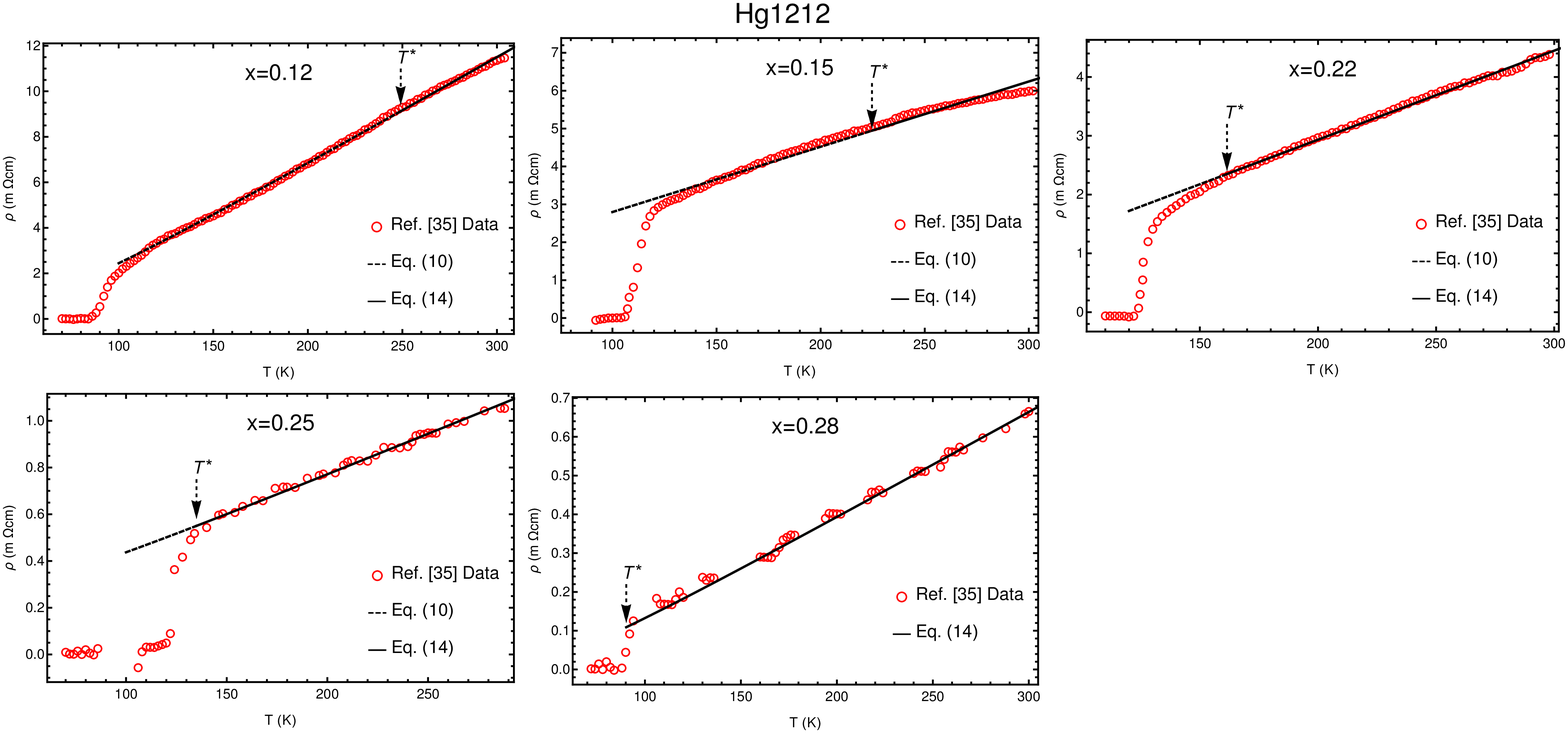}
		\caption{Comparison between the experimental data extracted from Ref.~\cite{yamamotopress} and the fitted Eq.~\eqref{eq_rho_SM} (with addition of $\rho_0$) for Hg1212.}
		\label{fig_res_Hg1212}
	\end{figure}
\end{widetext}

\end{document}